# Magnetic dielectric - graphene - ferroelectric system as a promising non-volatile device for modern spintronics

*Anatolii I. Kurchak[1], Anna N. Morozovska[2, 3*], and Maksym V. Strikha[1, 4]*

[1] *V.Lashkariov Institute of Semiconductor Physics, National Academy of Sciences of Ukraine, pr. Nauky 41, 03028 Kyiv, Ukraine*

[2] *Institute of Physics, National Academy of Sciences of Ukraine, pr. Nauky 46, 03028 Kyiv, Ukraine*

[3] *Bogolyubov Institute for Theoretical Physics, National Academy of Sciences of Ukraine, 14-b Metrolohichna str. 03680 Kyiv, Ukraine*

[4] *Taras Shevchenko Kyiv National University, Faculty of Radiophysics, Electronics and Computer Systems, pr. Akademika Hlushkova 4g, 03022 Kyiv, Ukraine*

The conductivity of the system magnetic dielectric (EuO) - graphene channel - ferroelectric substrate was considered. The magnetic dielectric locally transforms the band spectrum of graphene by inducing an energy gap in it and making it spin-asymmetric with respect to the free electrons. The range of spontaneous polarization (2 - 5)mC/m$^2$ that can be easily realized in thin films of proper and incipient ferroelectrics, was under examination. It was demonstrated, that if the Fermi level in the graphene channel belongs to energy intervals where the graphene band spectrum, modified by EuO, becomes sharply spin-asymmetric, such a device can be an ideal non-volatile spin filter. The practical application of the system under consideration would be restricted by a low Curie temperature of EuO,. However, alternative magnetic insulators with high Curie temperature (e.g. $Y_3Fe_5O_{12}$) can be used for a system operating under ambient conditions. Controlling of the Fermi level (e.g. by temperature that changes ferroelectric polarization) can convert a spin filter to a spin valve.

---

[*] Corresponding author. E-mail: anna.n.morozovska@gmail.com

# I. INTRODUCTION

Shortly after a field transistor with a graphene channel on a dielectric substrate was created in 2004 for the first time [1], multiple attempts have been made to use the unique properties of the new 2D-material in spintronics. At first graphene was proposed to be used as a non-magnetic spacer connecting two ferromagnetic contacts of the spin valve [2]. It has been experimentally shown that, due to the small spin-orbital interaction in graphene, the spin-relaxation length of a spin-polarized current at room temperature can be of 2 μm order (see, e.g., [3] and refs therein). However, at the same time, it was concluded that the graphene is poorly attractive for spintronics, since the magnetoresistance of the valve, $MR = \dfrac{R_{AP} - R_P}{R_P}$, is small due to the small number of conductance modes corresponding to the graphene channel Fermi energy in comparison with the analogous number of modes in ferromagnetic contacts ($R_{AP}$ and $R_P$ are the valve resistances for antiparallel and parallel orientation of magnetization at ferromagnetic contacts, respectively, see e.g. [2]). The physical consequences of the inequalities are explained in detail in Ref.[2].

Despite of pessimistic expectations effective spin valves with a graphene "spacer" and cobalt contacts have been created soon [4]. The spin valves operated at a sufficiently high 2D-concentration of carriers in the graphene channel ($n \approx 3.5 \times 10^{16}$ m$^{-2}$), and at high electric fields (~ ±70 kV/m) between the cobalt polarizer (source) and the analyzer (drain); and ~50% modulation of a useful signal was obtained. Since then, intensive efforts have been made to improve such valves. In particular, quite recently, the spin valve with cobalt contacts and 6 μm graphene channel have been created [5], at that spin polarization of the injection contact can be controlled by the bias and gate voltages. A detailed first principle theoretical study of the operation mechanism of a similar nano-valve with nickel contacts has been carried out in Ref.[6].

In parallel, an effective device has been proposed, which is either a spin valve or a spin filter, and no longer uses graphene as a nonmagnetic spacer, but as an active ferromagnetic element [7]. To realize this, a ferromagnetic dielectric EuO is imposed on the part of the graphene channel, which results in the strong spin polarization of the π-orbitals of graphene. As a result, the splitting of the graphene band states into the subbands with the orientation of the spin values "up" and "down" occurs, and also EuO induces the energy gap between these bands [8]. The transition between the states of the filter and the valve in [7] was induced by voltage at the lower gate.

In the series of works [9, 10, 11, 12, 13, 14] it is demonstrated how the ferroelectric substrate (instead of a usual dielectric) can be used for the doping of a graphene conductive channel by a significant number of carriers without the traditional application of the gate voltage. Thus, the concentration at which the observed effect [4] was revealed can be achieved by placing the graphene layer on a ferroelectric substrate with a "weak" spontaneous polarization of 5 mC/m² order that is only twice as high as the spontaneous polarization of Rochelle salt.

In this paper, we will show that it is possible to create a non-volatile spin valve / filter similar to that proposed in Ref.[7], where, however, the appropriate location of the Fermi level is provided not by the gate voltage, but by the spontaneous polarization of the ferroelectric substrate.

## II. THEORETICAL MODEL

The system geometry is depicted in **Fig. 1**. The single-layer graphene channel is considered as an infinitely thin two-dimensional (2D) gapless semiconductor of rectangular shape with length $L$ and width $W$. We regard that the graphene channel of length $L$ that is less than the free path of the electron $l$. Therefore the conductivity takes place in the ballistic regime. The graphene channel is deposited on a polarized single-domain ferroelectric film with a spontaneous polarization $P_S$. $P_S$ value determines the carrier concentration per unit area of the channel, $n = P_S/e$, where $e$ is the electron charge. We regard that the sign "+" of $P_S$ corresponds to a positive bound charge at the graphene-ferroelectric interface, and thus the graphene doping with electrons. Instead, the sign "-" corresponds to the negative bound charge at the interface and to the channel doping with holes. As in Ref.[7], a magnetic dielectric EuO with a length of $l \ll L$ is superimposed on the graphene channel. A top gate is imposed over the magnetic dielectric.

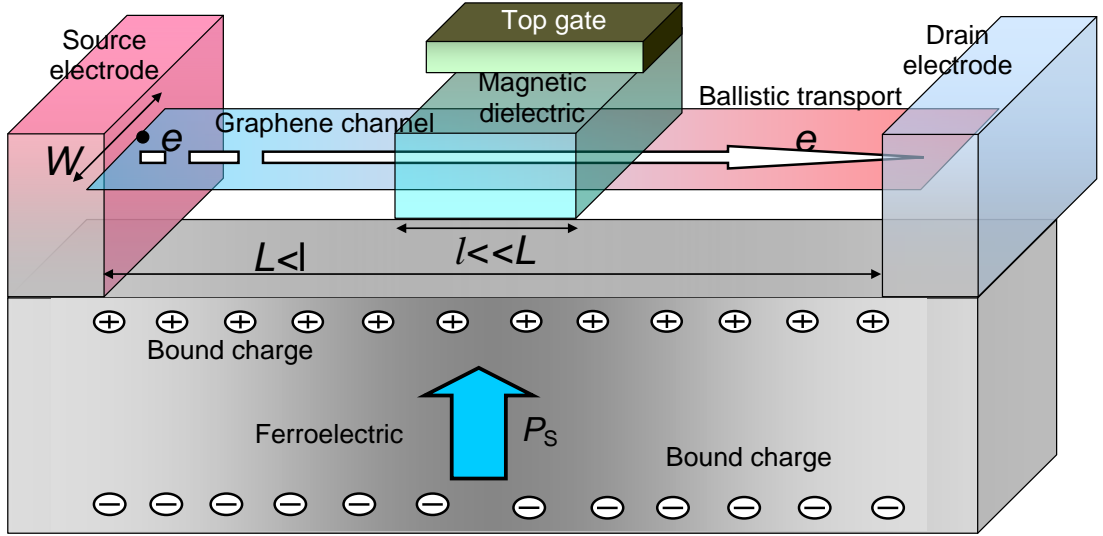

**Fig.1** Graphene single-layer of length $L$ and width $W$ is placed between magnetic dielectric (e.g. EuO) and ferroelectric film with wide domains. A top gate is imposed over the magnetic dielectric. The spontaneous polarization is $+P_S$ corresponding to the positive bound charge at the graphene-ferroelectric interface.

The single-layer graphene is a 2D gapless semiconductor with the linear band spectrum near the Dirac point [2]:

$$E_{\pm}(k) = \pm \hbar v_F k, \qquad (1)$$

where $k = \sqrt{k_x^2 + k_y^2}$ is the wave vector value, $v_F = 10^6$ m/s is a Fermi velocity, and the signs "+" and "-" correspond to the conduction and valence bands, respectively. In the graphene channel section located under a magnetic dielectric the spectra (1) undergoes modifications [8]:

$$E_s(k) = D_s \pm \sqrt{(\hbar v_s k)^2 + (\mathsf{D}_s/2)^2} \qquad (2)$$

Here the icon $s$ designates the two values of the spin projection ($s = \text{-},\bar{\ }$); the energies of the Dirac point shift are $D_{\text{-}} = 31$ meV and $D_{\bar{\ }} = -31$ meV, respectively. The energies determining the splitting are $\mathsf{D}_{\text{-}} = 134$ meV and $\mathsf{D}_{\bar{\ }} = 98$ meV; and the renormalized values of the electron velocity are equal to $v_{\text{-}} = 1.15 \times 10^6$ m/s and $v_{\bar{\ }} = 1.40 \times 10^6$ m/s, respectively. Following Ref.[7], the characteristic energies of the band edges arise in the band spectrum (2):

$$D_{\text{-}} + \mathsf{D}_{\text{-}}/2 = 98\,meV, \quad D_{\bar{\ }} + \mathsf{D}_{\bar{\ }}/2 = 18\,meV, \quad D_{\text{-}} - \mathsf{D}_{\text{-}}/2 = -36\,meV, \quad D_{\bar{\ }} - \mathsf{D}_{\bar{\ }}/2 = -80\,meV. \qquad (3)$$

The gap in the energy spectrum of graphene induced by a magnetic dielectric is shown schematically in **Fig. 2**. Note here, that first principle calculations [8] demonstrate strong doping of graphene by electrons from EuO, leading to the 1.4 eV shift of Fermi level in graphene on EuO into conduction band. Therefore a top gate is imposed over the magnetic dielectric in order to compensate by a proper applied gate voltage a local doping of graphene channel by electrons from proximity effect of EuO. Moreover, we treat the case of the short graphene on EuO section $l \ll L$. Therefore the doping of graphene conducting channel along all it's length $L$ is determined by the ferroelectric spontaneous polarization. This means that the position of the Dirac point "0" in a nonmagnetic graphene channel is twice closer to the edge of the upper band than to the lower one [see Eq.(3)], and so there is a certain asymmetry in the spin-polarized spectra under EuO for the ferroelectric polarization up and down directions.

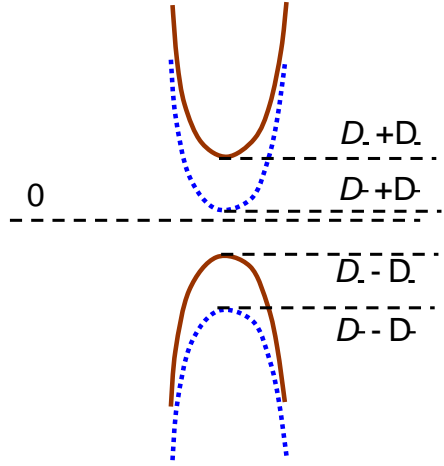

**Fig.2** The gap in the energy spectrum of graphene induced by a magnetic dielectric (adapted from Ref.[7]). Dotted curves correspond to the "down" spin states, solid curves are for states with spin "up". Zero energy "0" corresponds the position of the Dirac point in a nonmagnetic graphene channel.

When the Fermi energy is inside a wide energy window between the first two energy values in Eq.(3), then graphene becomes completely spin-polarized, i.e. 100% of carriers taking part in electro-transport have "down" spins. When the Fermi energy is in the window between the last two energy values in Eq.(3), the graphene is also completely spin-polarized, i.e. all conductivity carriers taking part in have "up" spins.

Our next assumption is that although the electron would pass through a non-magnetic graphene channel of length $L$ without dispersion, however, the presence of the section with length $l \ll L$ in the channel, where the graphene has pronounced magnetic properties, leads to the necessity to account for the local scattering of carriers. The intensity of the scattering depends on the carrier spin signs. Thus, the full conductivity of the graphene channel, taking into account the double degeneration of graphene at points $K, K'$, will be described by the modified Landauer formula (see, e.g., Refs.[15, 16]):

$$G = \sum_s G_s, \qquad G_s = \frac{2e^2}{\pi \hbar} M(E_F) T_s(E_F). \qquad (4)$$

Here $M(E_F)$ is the number of conductance modes, $T_s(E_F)$ is the transmission coefficient (in fact, the probability that the electron will pass without a scattering the "magnetic" section of length $l$). Both values correspond to the Fermi energy, and $T_s(E_F)$ depends also on the value of the electron spin, so for the full conductivity it is necessary to sum for both spin values.

Taking into account the relation $l \ll L$ we can assume that $M(E_F) = 0$ with high accuracy, when Fermi energy level is inside the energy gap of the spectrum (2), and lies between the 2-nd and 3-rd energies in Eq.(3). Outside the gap $M(E_F)$ is described by the expression [16],

$$M(E_F) = Int\left[\frac{2W}{\lambda_{DB}}\right],$$ and has the physical meaning of the number of de Broglie half-wavelengths $\lambda_{DB}/2$ for an electron in the graphene channel, which can be located at the width of this channel $W$. Symbol "Int" denotes the integer part.

Taking into account the known relation between 2D-concentration of electrons and Fermi energy in graphene,

$$E_F = \hbar v_F \sqrt{\pi n} = \hbar v_F \sqrt{\frac{\pi P_S}{e}}, \qquad (5)$$

we obtain for the de Broglie wavelength:

$$\lambda_{DB} = \frac{2\pi \hbar v_F}{E_F} = 2\sqrt{\frac{\pi e}{P_S}}. \qquad (6)$$

From Eq.(5), the possible range of Fermi energy tuning by the spontaneous polarization varying in the range $P_S \approx (0.01 - 1)$ C/m² is $(0.3 - 3.0)$ eV (for $v_F \approx 10^6$ m/s and $\hbar = 6.583 \times 10^{-16}$ eV·s).

In the case when the Fermi level lies outside the gap of the spectrum (2), the number of conductance modes is:

$$M(E_F) = Int\left[W\sqrt{\frac{P_S}{\pi e}}\right] \qquad (7)$$

Note that the Fermi energy of the considered system is related to the ferroelectric polarization by relation (5). Thus, for sufficiently high $P_S$, the conductivity (4) will depend on $P_S$ as $\sqrt{P_S}$, which is a direct consequence of the assumption of the ballistic nature of conductivity in the graphene channel [16].

Finally let's analyze the transmission coefficient $T_s(E_F)$. It was shown in Ref.[7] that $T_\uparrow(E_F) \approx 0$ and $T_\downarrow(E_F) \approx 1$ when the Fermi level is within the energy interval between the first and the second energy levels in Eq.(3). The standard for spintronics situation corresponds to the intensive scattering of electrons from the spin minority and to a much weaker scattering of electrons from the spin majority [2]. When the Fermi level falls into the interval between the third and the fourth energy values in Eq.(3), obviously we have an inverse case: $T_\uparrow(E_F) \approx 1$ and $T_\downarrow(E_F) \approx 0$. Finally, when the values of polarization are so high that the Fermi level lies either above the first or below the fourth energy levels in Eq.(4), both transmission coefficients are close to unity.

## III. DISCUSSION

As one can see from the analysis above, there are intervals of ferroelectric polarization, for which the considered system is an ideal spin filter, and the value of the ratio $\frac{G_\uparrow - G_\downarrow}{G_\uparrow + G_\downarrow}$ is close to 1. As the estimates show, the energies given by Eq.(3) correspond to sufficiently small polarizations of the order of mC/m$^2$, that is less than the spontaneous polarization of Rochelle salt.

Within continuous media Landau-Ginzburg-Devonshire approach [17], the value of the spontaneous polarization $P_S$ in thin ferroelectric films can be controlled by size effect, temperature $T$ and/or misfit strain $u_m$ originated from the film-substrate lattice constants mismatch [18, 19]. In particular, $P_S$ depends on $T$, $u_m$ and film thickness $h$ in the following way [17]:

$$P_S(T, u_m, h) = P_0 \sqrt{1 - \frac{T}{T_{cr}(u_m, h)}}, \qquad (8)$$

The spontaneous polarization $P_0$ corresponds to the bulk sample at zero Kelvin; and the temperature $T_{cr}(u_m, h)$ of the second order phase transition of the film to the paraelectric state has the form [17]

$$T_{cr}(u_m, h) \approx T_C^f \left(1 + \frac{2Q_{12} u_m}{\alpha_T T_C (s_{11} + s_{12})}\right) - \frac{l_S}{\alpha_T \varepsilon_0 \varepsilon_b^f (h + l_S)} \qquad (9)$$

Here, $T_C^f$ is the Curie temperature of bulk ferroelectric, $Q_{12}$ is the component of the electrostriction tensor, $s_{ij}$ are elastic compliances. The positive coefficient $\alpha_T$ is proportional to the inverse Curie-Weiss constant. Since the inequality $s_{11} + s_{12} > 0$ is valid for all ferroelectrics, the positive term $Q_{12} u_m$ increases $T_{cr}(u_m, h)$ and the negative term $Q_{12} u_m$ decreases it. The change of $T_C^f$ can reach several hundred Kelvins [18-17]. The graphene screening length $l_S$ is usually smaller (or significantly smaller) than 0.1 nm [20, 21]. $\varepsilon_b^f$ is an effective dielectric permittivity of interfacial or "passive" layer on ferroelectric surface [14]. Equation (9) is valid under the condition of an ideal electric contact between the ferroelectric film and graphene channel, and the absence of dead layer is assumed.

Within continuous media Landau approach [18] the expression for $P_0(T)$ follows from the minimization of Landau energy, $G_{Landau} = \int \left[ \alpha_T [T - T_C^f] \frac{P^2}{2} + \beta(T) \frac{P^4}{4} + \gamma(T) \frac{P^6}{6} \right]$ and has the form $P_0(T) = \sqrt{\frac{1}{2\gamma(T)}\left(\sqrt{\beta^2(T) + 4\gamma(T)\alpha_T(T_C^f - T)} - \beta(T)\right)}$. The expression for $P_0(T)$ and Eqs.(8)-(9) allow to vary $P_S$ from the bulk values of (0.5 – 0.05) C/m$^2$ order to the values hundreds of times smaller. For numerical calculations we chose the first order ferroelectric with relatively low $T_C^f$, namely BaTiO$_3$, which parameters a, b and g were obtained from Refs.[22, 23, 24, 25, 26] and are listed in **Table I**.

**Figure 3(a)** illustrates the situation for thin BaTiO$_3$ film on tensile substrate inducing misfit strains $u_m$ » 0.7%. The disadvantage of this choice is the fact that the tensile strain can result in the in-plane rotation of the polarization. The compressive strain enhances out-of-plane polarization and suppresses its rotation, but its value becomes substantially greater than 0.1 C/m$^2$ at a thickness slightly more than critical [see **Fig.3(b)**]. As a rule, $P_S$ increases with thickness and saturates for thin ferroelectric films in the absence of self-polarizing substrate or defects.

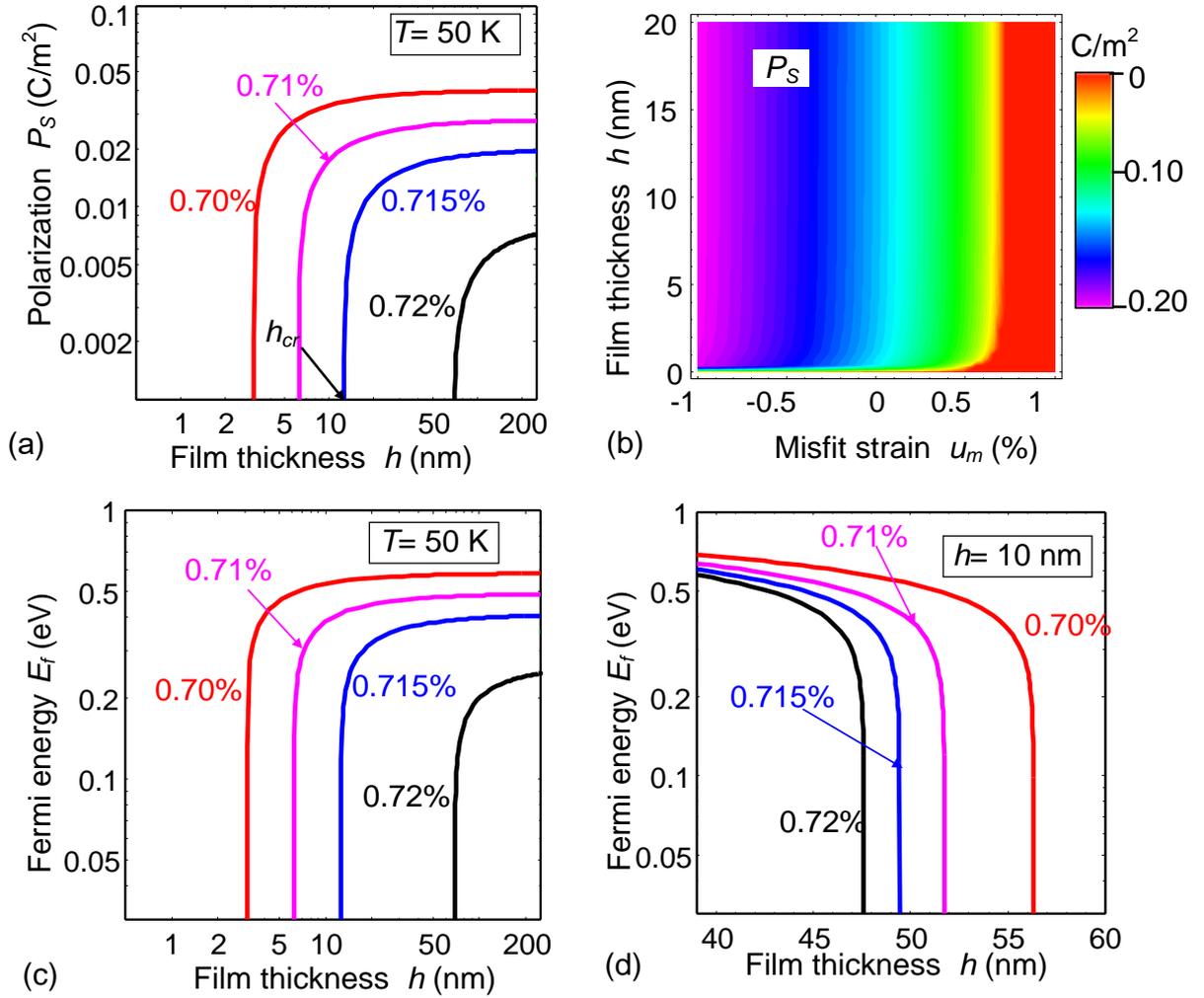

**Fig. 3.** (a) The dependence of the spontaneous polarization $P_S$ of a thin BaTiO$_3$ film on its thickness $h$ calculated at low temperature for different mismatch strains $u_m$ » 0.7% (red curve), 0.71% (magenta curve), 0.715% (blue curve), and 0.72% (black curve). (b) The dependence of $P_S$ on $h$ and $u_m$, calculated at low temperature. The dependences of the Fermi energy $E_F(T, u_m, h)$ on $h$ (c) and $T$ (d) calculated for $u_m$ » 0.7% (red curve), 0.71% (magenta curve), 0.715% (blue curve), and 0.72% (black curve). BaTiO$_3$ parameters are listed in **Table I**, $l_S$ =0.1 nm.

**Table I.** LGD parameters for bulk ferroelectric $BaTiO_3$

| $\varepsilon_b^f$ | $\alpha_T$ (C$^{-2}$·m J/K) | $T_C^f$ (K) | $\beta(T)$ (C$^{-4}$·m$^5$J) | $\gamma(T)$ (C$^{-6}$·m$^9$J) | $Q_{ij}$ (C$^{-2}$·m$^4$) | $s_{ij}$ ($\times 10^{-12}$ Pa$^{-1}$) |
|---|---|---|---|---|---|---|
| 120 | 6.68×10$^5$ | 381 | $\beta_T(T-393)$–8.08×10$^8$ $\beta_T$=18.76×10$^6$ | $\gamma_T(T-393)$+16.56×10$^9$ $\gamma_T$= -33.12×10$^7$ * | $Q_{11}$=0.11, $Q_{12}$= -0.043, $Q_{44}$=0.059$^0$ | $s_{11}$=8.3, $s_{12}$= -2.7, $s_{44}$=9.24 |

*These parameters are valid until $\gamma$>>0, i.e. for T<400 K.

Controlling of the Fermi level (e.g. by temperature changes of the ferroelectric polarization) can convert a spin filter to a spin valve. Actually, $E_F \sim \sqrt{P_S}$ accordingly to Eq.(5), and the temperature dependence of polarization is given by Eq.(8). From these equations,

$$E_F(T, u_m, h) = E_F^0 \left(1 - \frac{T}{T_{cr}(u_m, h)}\right)^{1/4}. \qquad (9)$$

Since the factor $E_F^0 = \hbar v_F \sqrt{\pi P_0/e}$ can vary in the range $(0.3 - 3.0)$ eV for different ferroelectric films with $P_0 \approx (0.01 - 1)$ C/m$^2$ at temperatures lower (or significantly lower) that $T_{cr}$, the value of Fermi energy in Eq.(9) decreases from $E_F^0$ to 0 with the temperature increase from 0 to $T_{cr}$. The influence of temperature and film thickness on Fermi energy is presented in **Figs 3(c)** and **3(d)**, respectively. From the figures, $E_F(T, u_m, h)$ varies in the range $(0.03 - 0.6)$ eV for the temperatures lower 50 K. Notably, that the actual temperature range, and so $T_{cr}(u_m, h)$, should be significantly smaller than the EuO's Curie temperature $T_C$=77 K.

In addition to the size and substrate effects in thin films, incipient ferroelectrics can be used to provide the required values of polarization ~mC/m$^2$. Their polarization is small, since it is a secondary, not a primary order parameter. In particular, thin films of binary oxides (titanium and hafnium oxides) [27, 28] can be suitable candidates, since they have a small spontaneous polarization that depends significantly on the film thickness and defect concentration, and varies in a wide range from 0.2 C/m$^2$ to 0.002 C/m$^2$, as well as ferromagnetic properties [29]. So they are multiferroics, which can combine the magnetic properties of EuO with low ferroelectric polarization [30].

In **Fig. 4**, the value of conductivity (4) is shown as a function of spontaneous polarization $P_S$ for different values of graphene channel width $W$ = 50, 100 nm and 200 nm. Mention that for the width values smaller than 50 nm additional quantization along $y$ axis should be taken into consideration, and Eqs. (2)-(4) and (7) should be corrected accurately. The difference in the form of the curves in **Fig.4** calculated for different values of $W$ is due to the increase in the number of conductance modes (7) occurring with increasing of $W$. Allowing for the assumption about the ballistic nature of transport in the graphene channel, the conductivity depends on the square root of the charge carriers concentration in the channel (see, e.g. [16], and, therefore, on the square root of the polarization). If, for sufficiently long channels, the conduction regime becomes diffusive, it will result to the additional factor $l(E_F)/L$ in expression (4) (see, e.g. [16]), where $l(E_F)$ is the electron free path corresponding to the Fermi level energy. Consequently, the dependence of the conductivity on $P_S$ depends on the dominant scattering mechanisms of carriers in the graphene channel. If the scattering occurs predominantly on ionized impurities in substrate (the most common case) then the conductivity will depend linearly on the carrier concentration and polarization [16].

**Figure 5(a)** shows the dependence of the ratio $\dfrac{G_\uparrow - G_\downarrow}{G_\uparrow + G_\downarrow}$ on the ferroelectric polarization calculated on the basis of Eqs.(4)-(8) and in accordance with **Fig 4**. The ratio is $W$-independent at $W \geq 50$ nm. Dependences of the normalized $G_\uparrow$ (left side) and $G_\downarrow$ (right side) on $W$ and $P_S$ are shown in **Fig.5(b)**. Semi-transparent rectangle indicates the region $W < 50$ nm, where the validity of our approach should be checked additionally due to reasons mentioned above. As it is evident from **Fig. 5** there is a range of $P_S$ from $10^{-4}$ C/m² to $1.2 \times 10^{-3}$ C/m² for direction $+P_S$, and from $10^{-4}$ C/m² to $0.8 \times 10^{-3}$ C/m² for direction $-P_S$, in which the proposed device is an ideal non-volatile spin filter. Such polarization values, as demonstrated by the analysis performed above, can be provided by e.g. thin films of BaTiO$_3$ with thickness less than 100 nm.

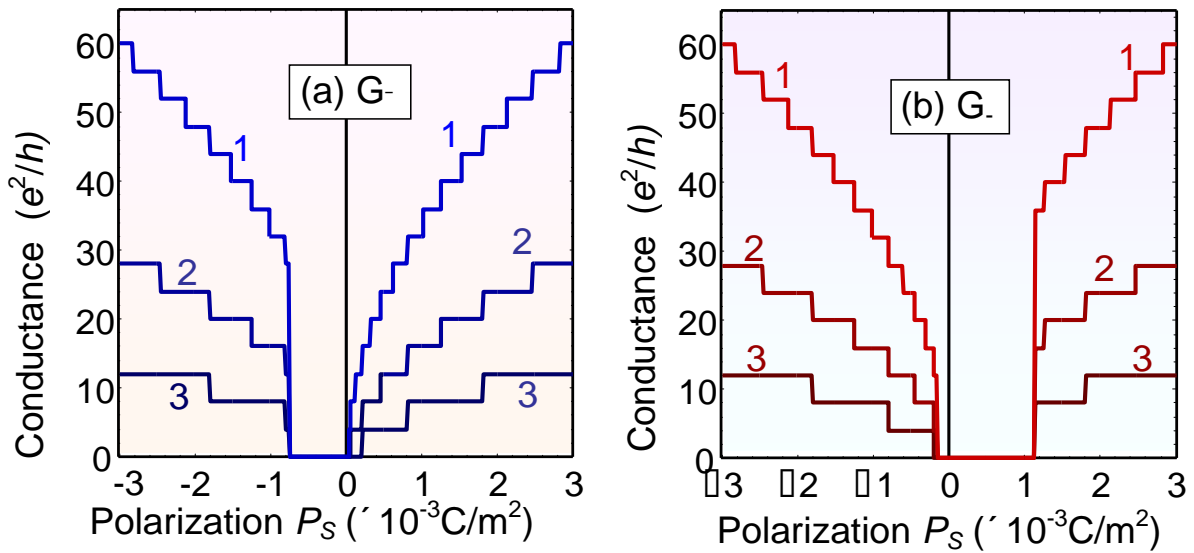

**Fig. 4.** The dependence of the graphene channel conductivity $G_\uparrow$ (a) and $G_\downarrow$ (b) on the spontaneous polarization $P_S$ calculated for several widths of the graphene channel $W = 50$, 100 and 200 nm (step-like curves 1, 2 and 3). The conductance is normalized on Klitzing constant $e^2/2\pi\hbar$.

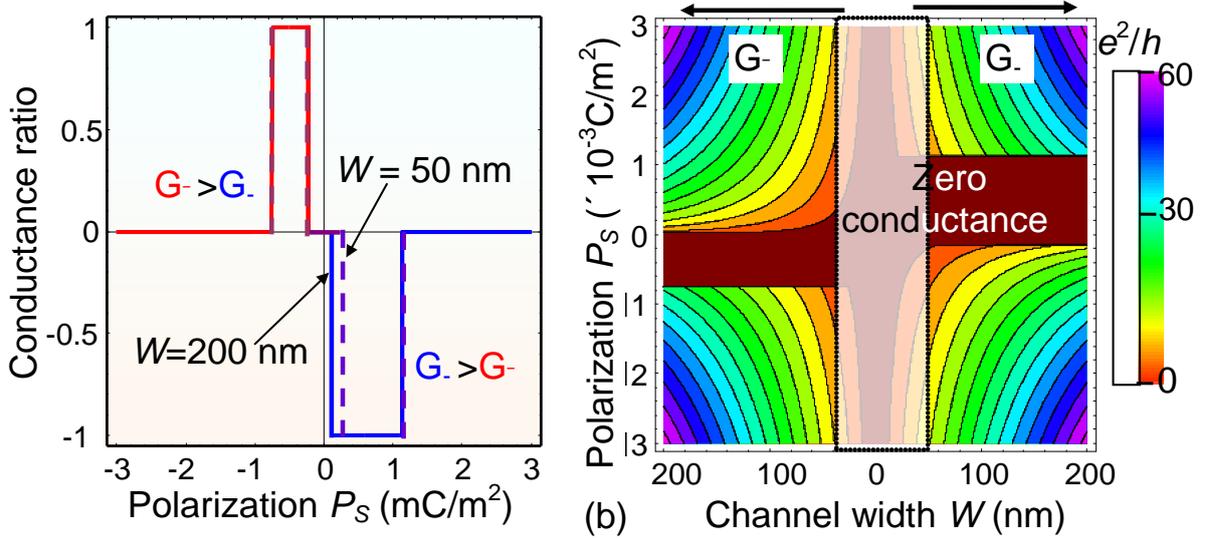

**Fig. 5. (a)** Dependence of the ratio $\dfrac{G_\uparrow - G_\downarrow}{G_\uparrow + G_\downarrow}$ on the ferroelectric polarization $P_S$ calculated for $W = 50$ nm (dashed curves) and 200 nm (solid curves). **(b)** Dependences of the normalized $G_\uparrow$ (left side) and $G_\downarrow$ (right side) on $W$ and $P_S$. The conductance is normalized on Klitzing constant $e^2/2\pi\hbar$. Semi-transparent rectangle indicates the region $W < 50$ nm, where our approach becomes invalid because of additional quantisation along $y$ axis.

## IV. CONCLUSION

We considered the conductivity of the magnetic dielectric system (EuO) placed on the graphene conductive channel, which in turn was deposited at the ferroelectric substrate. In this case, the magnetic dielectric locally transforms the band spectrum of graphene by inducing an energy gap in it and making it asymmetric with respect to the spin of the free electrons. The range of spontaneous polarization of ferroelectrics $(2 \div 5)$ mC/m$^2$, which can be easily realized in thin $(10 - 100)$ nm films of proper and incipient ferroelectrics, was under examination. It was demonstrated, that if the Fermi level in the graphene channel belongs to energy intervals where the graphene band spectrum, modified by EuO, becomes sharply spin-asymmetric, such a device can be an ideal non-volatile spin filter.

Controlling of the Fermi level (e.g. by temperature changes of the ferroelectric polarization) can convert a spin filter to a spin valve. Actually, $E_F(T, u_m, h) \approx E_F^0 \left(1 - \dfrac{T}{T_{cr}}\right)^{1/4}$ accordingly to Eq.(9), and it can vary in the range $(0.03 \div 0.6)$ eV for graphene on BaTiO$_3$ at temperatures lower 50 K. Note that except for relatively low ferroelectric transition temperature $T_C^f$, there is no particular reason to for choosing BaTiO$_3$ instead of e.g. Pb$_x$Zr$_{1-x}$TiO$_3$ for some composition x near the morphotropic phase boundary where both $T_C^f$ and $P_S$ can be relatively small. Derived analytical expressions allow calculations for any ferroelectric substrate that can be of potential interest for concrete experiment. We hope that our theoretical predictions will stimulate new experiments for different heterostructures top gate/ferromagnetic/graphene/ferroelectric substrate.

Note, that the problem solved above has a framework character. The practical application of the system under consideration is restricted by a relatively low ferromagnetic transition temperature of EuO, $T_C^m = 77K$. However, as it was demonstrated by Hallal et al. [31], alternative magnetic insulators with higher Curie temperatures can cause similar local transformation of graphene band spectrum. According to the first principles calculations [31], energy gaps, imposed in graphene by magnetic insulator $Y_3Fe_5O_{12}$ (YIC), are similar to the ones described by Eq.(3); however, the "useful" energy ranges with spin asymmetry are several times wider there, which makes such a system more convenient for practical usage. High ferromagnetic transition temperature of YIC ($T_C^m = 550K$) permits the system to operate under ambient conditions. However, quite recently Song [32] demonstrated that the gate-induced spin valve based on graphene/YIG (or on graphene/EuS) also induces a heavy electron doping, 0.78eV, which corresponds to a giant spontaneous polarization of 1.5 $C/m^2$. Thus the device we proposed cannot operate without the top gate.

Complementary to the said above, consideration of the graphene on a ferroelectric-ferromagnetic multiferroic that simultaneously provides both the magnetic asymmetry of the graphene spectrum and its doping with free carriers can be challenging for advanced applications in spintronics. We hope to consider the problem in future.

**Authors' contribution.** A.I.K. wrote the codes for results visualization. A.N.M. performed analytical calculations in section III and generated plots. M.V.S. generated the research idea, performed analytical calculations in section II and wrote the manuscript draft. All authors contribute to the results discussion and manuscript improvement.

A.I.K. and A.N.M. acknowledges the State Fund for Fundamental Research (grant number F81/41481). A.N.M. work was partially supported by the National Academy of Sciences of Ukraine (projects No. 0118U003375 and № 0118U003535).